\documentclass[a4paper,11pt]{article}
\usepackage[top=2.54cm,bottom=2.54cm, left=2.54cm, right=2.54cm]{geometry}

\usepackage{amsmath}
\usepackage{amsfonts}
\usepackage{amssymb}
\usepackage{graphicx}
\usepackage{bm}
\usepackage{enumerate}
\usepackage{color}
\usepackage[sort,comma]{natbib}
\usepackage{float}
\usepackage[colorlinks=true, allcolors=blue]{hyperref}
\usepackage{changepage}
\usepackage{dirtytalk}
\usepackage{pdflscape}
\usepackage{booktabs}

\title{Fast and light-weight energy statistics using the \textsf{R} package \texttt{estats}}
\author{Michail Tsagris$^1$ and Manos Papadakis$^2$\\
$^1$ Department of Economics, University of Crete, Rethimnon, Greece  \href{mailto:mtsagris@uoc.gr}{mtsagris@uoc.gr} \\
$^2$ Independent Researcher, Heraklion, Greece  \href{mailto:papadakm95@gmail.com}{papadakm95@gmail.com} 
}

\begin{document}

\maketitle

\begin{center}
\textbf{Abstract}
\end{center}

$\{\text{Fast and memory-efficient computation of energy related statistical quantities.}\}$ \\
\\
Energy statistics ($\mathcal{\varepsilon}$--statistics) enable powerful non-linear dependence measures such as distance correlation, but their computational burden has limited application to large datasets. We present memory-efficient algorithms that compute $\mathcal{\varepsilon}$--statistics related quantities by calculating pairwise distances on-the-fly rather than storing full distance matrices. Our methods achieve 5-156$\times$ speed improvements over existing implementations while reducing memory requirements from $O(n^2)$ to $O(n)$. These advances enable energy statistics computation with sample sizes exceeding tens of thousands observations—previously infeasible with standard implementations—facilitating their use in modern applications across statistics, bioinformatics, and machine learning where large-scale datasets are frequently met. The following cases are demonstrated: energy distance, univariate and multivariate distance variance, distance covariance, (partial) distance correlation and hypothesis testing for the equality of univariate distributions. Functions to compute the aforementioned energy statistics, among others, are available in the \textit{R} package \textsf{estats}. 
\\
MSC: 6208, 6204 

\section{Introduction}
\cite{szekely2007,szekely2009} pioneered the concept of $\mathcal{\varepsilon}$--statistics. In their seminal works, they introduced the distance correlation and covariance, measures of non-linear correlation and, in essence, dependence between two random variables, in arbitrary dimensions. These two primary works have received thousands of citations, arguing that distance correlation is widely employed and has evolved as a classic measure. Distance correlation has been extensively applied across multiple domains, such as variable selection \citep{li2012}, network analysis \citep{maccarron2023}, and time series \citep{edelmann2019}. Applications of distance correlation extend beyond statistics to include finance \citep{ugwu2023}, bioinformatics \citep{hou2022}, and physics \citep{kasieczka2020} inter alia. 

However, the computational burden associated with this quantity constitutes a significant limitation to the widespread applicability of this novel correlation, which was highly praised, and characterised as the correlation of the 21st century, by \cite{speed2011}. Consider a bioinformatics researcher analyzing gene expression data with tens of thousands of multivariate observations. Computing distance correlation (and related quantities) using existing methods would require a computer with high capabilities. Large sample sizes render distance correlation computationally intractable using existing implementations.

To mitigate the computational cost associated with the distance correlation \cite{huo2016} proposed a fast implementation of the distance correlation, but for the univariate case. Their algorithm possesses a computational complexity of $O(n \log{n})$, which is the same as required for the computation of the Spearman's correlation, allowing for the computation of distance correlation even with millions of observations. \cite{chaudhuri2019} provided a different algorithm that improved the computational time of the dyadic algorithm of \cite{huo2016}, still for the case of vectors. For the case of two or more dimensions, the current implementation in the \textit{R} package package \textsf{energy} \citep{energy2024} is limited by the number of observations (or sample size), and this limitation applies to the partial distance correlation, distance variance, distance covariance, energy distance, and the equality of distributions test. The \textit{R} package \textsf{dcortools} \citep{dcortools2022} on the contrary does not have this limitation, as it is memory efficient\footnote{Note that the package \textsf{dcortools} maintains the same memory requirements as our implementations.}. 

\cite{huang2022} pursued an alternative approach and proposed an approximate distance covariance estimator that is based upon random projections. The randomness of their algorithm vanishes as the sample size tends to infinity, thereby making it suitable for large-scale dataset. However, there is no implementation of their algorithm in any \textit{R} package. 

The computational burden, both in time and memory remains, regardless of the dimensionality of the data, still remains for $\varepsilon$-statistics and the current paper comes to adress this issue. Our contributions are as follows. We introduce methods and techniques for simultaneously mitigating the computational complexity and memory demands associated with the computation of certain $\varepsilon$--statistics. In the multivariate setting, we adopt the formulas proposed by \cite{szekely2023}, whereas in the univariate case, we leverage established mathematical identities. These methodological advancements play a crucial role in facilitating the efficient calculation of various $\varepsilon$--statistics, including the energy distance, distance variance, covariance, correlation and partial distance correlation, and hypothesis testing for the equality of univariate distributions. We also implemented, in \textit{R}, Huang and Huo's method \citep{huang2022} for computing an approximate distance covariance and compared it to our implementation. The relevant functions to compute the $\varepsilon$--statistics exist in the \textit{R} package \textsf{estats} \citep{estats2025}. The \textsf{estats} package is dedicated to energy statistics, works as a wrapper package that collects functions from other packages for completeness purposes. It imports the packages \textsf{Rfast} \citep{rfast2025} and \textsf{Rfast2} \citep{rfast22025} which contain implementations either in \textit{C++}, with the exception of the test for equal distributions that is implemented in \textit{R}. The \textsf{estats} package also imports functions from the packages \textsf{dcov}\footnote{This is for the distance correlation between two univariate variables, fully implemented in \textit{C++}.} \citep{dcov2020} and \textsf{pdcor}\footnote{This is for the partial distance correlation, that partially uses \textit{R} calling \textit{C++} functions.} \citep{pdcor2025}.

The next section briefly describes the $\varepsilon$--statistics showcasing their computationally efficient implementations, while Section \ref{secrfast} presents the relevant functions in the \textit{R} package \textsf{estats}. Section \ref{seccost} compares the efficiency of the functions in \textsf{estats} to packages \textsf{energy} and \textsf{dcortools}, and finally Section \ref{secconc} concludes the paper.

\section{$\varepsilon$--statistics} \label{secenergy}
Assume we have two (multivariate) variables $\bm{X}, \bm{Y} \in \mathbb{R}^p$, with cumulative distribution functions $F$ and $G$ respectively.

\subsection{Energy distance}
The energy distance between these two distributions is 
\begin{eqnarray*} 
\mathcal{\varepsilon}\left(F, G \right)=2E\left|\bm{X}-\bm{Y}\right| - E\left|\bm{X}-\bm{X}'\right| - E\left|\bm{Y}-\bm{Y}'\right|,
\end{eqnarray*}
where $\bm{X}'$ is an independent and identically distributed (i.i.d.) copy of $\bm{X}$, and $\bm{Y}'$ is an i.i.d. copy of $\bm{Y}$. 

The sample version of the energy distance is 
\begin{eqnarray} \label{energy}
\mathcal{\varepsilon}_n\left(\bm{X}, \bm{Y} \right)=\frac{2}{nm}\sum_{i=1}^n\sum_{j=1}^m\|\bm{X}_i-\bm{Y}_j \| - 
\frac{1}{n^2}\sum_{i=1}^n\sum_{j=1}^n\|\bm{X}_i-\bm{X}_j \| -
\frac{1}{m^2}\sum_{i=1}^m\sum_{j=1}^m\|\bm{Y}_i-\bm{Y}_j \|,
\end{eqnarray}
where $\| \cdot \|$ denotes the Euclidean norm, $\bm{X}_i$ and $\bm{Y}_j$ refer to the $i$-th and $j$-th observations of $\bm{X}$ and $\bm{Y}$, respectively, and $n$ and $m$ are their corresponding sample sizes. 

\subsection{Distance variance, covariance and correlation}
For the next three cases under consideration denote by $A$ and $B$ the Euclidean distance matrices of $\bm{X}$ and $\bm{Y}$, respectively, with $n=m$, and $a_{ij}=\|\bm{X}_i-\bm{X}_j \|$ and $b_{ij}=\|\bm{Y}_i-\bm{Y}_j \|$ denote the $(i,j)$ elements of these matrices. We next define the doubly centered matrix $\widetilde{A}$ whose entries are 
\begin{eqnarray*}
\widetilde{A}_{ij}= a_{ij}-\bar{a}_{i.}-\bar{a}_{.j}+\bar{a}_{..} \ \ (i,j=1,\ldots,n),
\end{eqnarray*}
where
\begin{eqnarray*}
\bar{a}_{i.}=\frac{1}{n}\sum_{j=1}^na_{ij}, \ \ \bar{a}_{.j}=\frac{1}{n}\sum_{i=1}^na_{ij} \ \ \text{and} \ \ \bar{a}_{..}=\frac{1}{n^2}\sum_{i,j=1}^na_{ij}.
\end{eqnarray*}
Similarly, the $(i,j)$ elements of the doubly centered matrix $\widetilde{B}$ are
\begin{eqnarray*}
\widetilde{B}_{ij}= b_{ij}-\bar{b}_{i.}-\bar{b}_{.j}+\bar{b}_{..} \ \ (i,j=1,\ldots,n),
\end{eqnarray*}

The sample distance covariance, variance, correlation and partial correlation are \citep{szekely2009,szekely2023} 
\begin{subequations}
\begin{eqnarray} 
\mathcal{V}_n^2\left(\bm{X},\bm{Y}\right)&=&\frac{1}{n^2}\sum_{i,j=1}^n\widetilde{A}_{ij}\widetilde{B}_{ij}  \label{dcov} \\
\mathcal{V}_n^2\left(\bm{X}\right)&=&\mathcal{V}_n^2\left(\bm{X},\bm{X}\right)=\frac{1}{n^2}\sum_{i,j=1}^n\widetilde{A}_{ij} \label{dvar} \\
\mathcal{R}_n^2\left(\bm{X},\bm{Y}\right)&=&\frac{\mathcal{V}_n^2\left(\bm{X},\bm{Y}\right)}{\sqrt{ \mathcal{V}_n^2\left(\bm{X}\right)\mathcal{V}_n^2\left(\bm{Y}\right)}} \label{dcor} \\
\mathcal{R}_n^{2}\left(\bm{X},\bm{Y}~|~\bm{Z}\right)&=&\frac{\mathcal{R}_n^{2}\left(\bm{X},\bm{Y}\right) - \mathcal{R}_n^{2}\left(\bm{X},\bm{Z}\right)\mathcal{R}_n^{2}\left(\bm{Y},\bm{Z}\right)}{\sqrt{1-\mathcal{R}_n^{2}\left(\bm{X},\bm{Z}\right)^2}\sqrt{1-\mathcal{R}_n^2\left(\bm{Y},\bm{Z}\right)^2}}. \label{pdcor}
\end{eqnarray}
\end{subequations}

The distance covariance (\ref{dcov}) using an alternative formula \citep{szekely2023} is computed as follows
\begin{eqnarray} \label{dcov2}
\mathcal{V}_n^2\left(\bm{X},\bm{Y}\right)=\frac{\sum_{1 \leq i\neq j \leq n}^na_{ij}b_{ij}}{n^2}-\frac{2\sum_{i=1}^na_{i.}b_{i.}}{n^3}+\frac{a_{..}b_{..}}{n^4}.
\end{eqnarray}

The bias-corrected distance covariance entails slightly different denominators \citep{szekely2023}
\begin{eqnarray} \label{bcdcov}
\mathcal{V}_n^{*2}\left(\bm{X},\bm{Y}\right)=\frac{\sum_{1\leq i\neq j \leq n}^na_{ij}b_{ij}}{n(n-3)}-\frac{2\sum_{i=1}^na_{i.}b_{i.}}{n(n-2)(n-3)}+\frac{a_{..}b_{..}}{n(n-1)(n-2)(n-3)}.
\end{eqnarray}

In the case of distance variance (\ref{dvar}) changes accordingly to become
\begin{eqnarray} \label{dvar2}
\mathcal{V}_n^2\left(\bm{X}\right)=\frac{\sum_{1 \leq i\neq j \leq n}^na_{ij}^2}{n^2}-\frac{2\sum_{i=1}^na_{i.}^2}{n^3}+\frac{a_{..}^2}{n^4}.
\end{eqnarray}
while the unbiased distance variance becomes
\begin{eqnarray} \label{bcddvar}
\mathcal{V}_n^{*2}\left(\bm{X}\right)=\frac{\sum_{1 \leq i\neq j \leq n}^na_{ij}^2}{n(n-3)}-\frac{2\sum_{i=1}^na_{i.}^2}{n(n-2)(n-3)}+\frac{a_{..}^2}{n(n-1)(n-2)(n-3)}
\end{eqnarray}
and the unbiased (partial) distance correlation changes accordingly.

\subsection{Approximate distance covariance}
\cite{huang2022} proposed an approximate distance covariance for matrices, based upon random projections. Their algorithm proceeds as follows
\begin{enumerate}
\item Let $C_p = \sqrt{\pi} \frac{\Gamma \left( (p + 1)/2 \right)}{\Gamma(p/2)}$ and $C_q = \sqrt{\pi} \frac{\Gamma( (q + 1)/2}{\Gamma(q/2)}$, where $p$ and $q$ denote the dimensionality of the matrix $\bm{X}$ and $\bm{Y}$, respectively. 
\item Draw two vectors $\bm{u}$ and $\bm{v}$ from the uniform distribution on the hyper-sphere with dimensions $p$ and $q$. 
\item Compute the univariate distance covariance using the fast method of \cite{huo2016} or of \cite{chaudhuri2019} on the projected vectors, $\Omega_n=C_p C_q \mathcal{V}_n^2\left(\bm{Xu},\bm{Yv}\right)$.
\item Repeat Steps 2-3 $K$ times and compute the average $\widetilde{\Omega}_{n,K}=\frac{1}{K}\sum_{k=1}^K\Omega_n^k$.
\end{enumerate}
To ensure approximation accuracy, a large value of $K$ is required. The authors suggested $K=50$, but in our simulation studies we explored $K=50$ and $K=100$. Since no \textit{R} implementation of the algorithm of \cite{chaudhuri2019} exists in an \textit{R} package we implemented the approximate distance covariance using the algorithm of \cite{huo2016} available in the \textit{R} package \textsf{dcov} \citep{dcov2020}.

\subsection{Testing for equality of two distributions}
To test for the equality of two univariate (and multivariate distributions), the test statistic is $T=\frac{nm}{n+m}\mathcal{\varepsilon}_n\left(\bm{X}, \bm{Y} \right)$. Permutations are used, where observations are randomly permuted between the two samples and Eq. (\ref{energy}) is computed. This process is repeated $B$ times, and the proportion of times the energy distance test statistics of the permuted data exceeds the energy distance test statistic computed at the observed data serves as an approximate p-value.

\section{Fast and light-weight computation of the $\varepsilon$--statistics}
The \textsf{R} package \textsf{energy} computes the distance (co)variance using Equations (\ref{dcov}) and (\ref{dvar}), respectively. However, this approach has two limitations, the computation and the storage of the distance matrix, thus requiring $O(n^2)$ memory. In the case of distance covariance or distance correlation, two matrices must be computed and stored. In scenarios where the sample size of observations approaches tens of thousands, it is imperative that the computer system possesses substantial memory capacity to facilitate the requisite computations. Alternatively, one may utilize available storage space on the hard drive, although this reduces memory usage, computational time remains substantial. A different approach involves computing the lower triangular matrix, which requires less memory, but the resulting reduction of the computational cost is only marginal. 

Utilization of formulas presented in Eq. (\ref{dcov2})-(\ref{bcddvar}) reduces the memory requirements to $O(n)$. For example, consider the case of two matrices having dimensions $n=20,000$ and $p=10$. Using \textsf{energy}, each distance matrix computation $A$ and $B$ requires $n^2 \times 8$ bytes (assuming double precision), thus $16 \times n^2=6.4$GB. The \textsf{estats} functions require $16 \times n + 16\times n\times p=$3.52MB, that is 0.055\% of the memory required by \textsf{energy}, or, a memory reduction of more than 1,800 times. 

\subsection{Computation of distance correlation and related functions}
\textsf{estats} computes the same quantities, distance covariance and variance, using Equations (\ref{dcov2}) and (\ref{dvar2}), respectively, but computes the required distances \textit{on-the-fly}, avoiding computation of the full distance matrix and minimizing the memory requirements. We will examine the computation of the distance variance (\ref{dvar2}) (evidently, the same approach applies to the bias-corrected versions). Equation (\ref{dvar2}) comprises three terms, all related to $a_{ij}$, that is, the Euclidean distance between the $i$-th and $j$-vector. The three quantities are 
\begin{eqnarray} \label{gammas}
\gamma_1=\sum_{1 \leq i\neq j \leq n}^na_{ij}^2, \ \ \gamma_2=2\sum_{i=1}^na_{i.}^2 \ \ \text{and} \ \ \gamma_3=a_{..}^2.
\end{eqnarray}

Using two nested \texttt{for()} loops one may compute all terms of Equation (\ref{dvar2}) at one pass. At iterations $i$ and $j$, the value of $a_{ij}$ is computed and summed yielding quantity $\gamma_3$. The values $a_{ij}$ are next summed in two places, one place related to quantity $\gamma_1$ and one place related to quantity $\gamma_2$. To compute the quantity $\gamma_2$ which must compute the sum of $a_{ij}$ with respect to $j$, thus, at iteration $i$ one must store the $n$-sized vector of the $a_{ij}$ values and then sum them to obtain $a_{i.}$. Then the $a_{i.}$ is squared and summed, contributing to the computation of quantity $\gamma_1$. The extra memory requirements, apart from the one required by the data sets themselves is the memory required by the $n$-sized vector of the $a_{ij}$ values. The same optimizations are employed for the cases of the distance covariance (\ref{dcov2}) and correlation (\ref{dcor}), but the computations for both datasets are performed concurrently.

Specifically for the univariate distance variance, the identity in the Gini coefficient \citep{esteban2004} facilitates fast computation of the second term $\gamma_2$, of Eq. (\ref{gammas}),
\begin{eqnarray*}
a_{i.} = (2i - n)x_{(i)} + \sum_{i=1}^nx_i - 2\sum_{k=1}^ix_{(k)},
\end{eqnarray*}
where $x_{(i)}$ denote the ordered values, in ascending order. The time complexity of the computation of $a_{i.}$ is $O(n \log{n})$, which is the complexity of sorting $n$ numbers\footnote{The time complexity of the sum computations is $O(n)$.}. To compute $\gamma_1$, we can apply the following identity
\begin{eqnarray*}
\sum_{i\neq j}^na_{ij}^2=\sum_{1 \leq i\neq j \leq n}^n|x_i-x_j|^2 = n \sum_{i=1}^nx_i^2 - \left(\sum_{i=1}^nx_i\right)^2.
\end{eqnarray*}
The last term in the above equation (the sum of all numbers), has already been computed (it is required in $\gamma_2$) and there is no need to compute it twice, thereby avoiding the redundant computation of the sum of all pairwise squared differences. 

Finally, for the partial distance correlation, \textsf{Rfast}'s relevant function, written in \textit{R}, employs the formula of Eq. (\ref{pdcor}) computing the necessary distance correlations rendering it computationally efficient. 

\subsection{Computation of energy distance}
The energy distance (\ref{energy}) employs a similar optimization strategy. The energy consists of three quantities, the sum of all pairwise distances between the matrix $\bm{X}$ and $\bm{Y}$, and the sum of all pairwise distances of $\bm{X}$ and $\bm{Y}$. The pairwise distances are computed again \textit{on-the-fly}, that is, each pairwise distance is computed and summed. This avoids computing, and storing, the whole distance matrix and then computing its sum. 

\subsection{Computations for testing the equality of two univariate distributions}
As mentioned earlier, random permutations occur repeatedly, and the constant factor $\frac{nm}{n+m}$ that multiplies the energy distance (\ref{energy}) does not influence the p-value, and can be omited. With $\bm{x}$ and $\bm{y}$ denoting two univariate random variables, with cumulative distribution functions $F$ and $G$, respectively, whose equality we wish to test, Eq. (\ref{energy}) consists of three terms
\begin{eqnarray*} 
\delta_1 = \sum_{i=1}^n\sum_{j=1}^m|x_i-y_j |, \ \ 
\delta_2 = \sum_{i=1}^n\sum_{j=1}^n|x_i-x_j | \ \ \text{and} \ \ 
\delta_3 = \sum_{i=1}^m\sum_{j=1}^m|y_i-y_j |.
\end{eqnarray*}

The use of the identity in the Gini coefficient \citep{esteban2004} ensures fast computation of the sum of all pairwise distances, 
\begin{eqnarray*} 
\delta_2 = \sum_{i=1}^n\sum_{j=1}^n|x_i-x_j | = 2 \sum_{i=1}^n(i x_{(i)} ) - (n + 1) \sum_{i=1}^nx_i.
\end{eqnarray*}
The same formula applies to the $\delta_3$ quantity. Finally, the quantity $\delta_1$ can be computed via the following identity
\begin{eqnarray*}
\delta_1= 2 \sum_{i=1}^{n+m}( i z_{(i)} ) - (n + m + 1)\sum_{i=1}^{n+m}z_i - \delta_2 - \delta_3,
\end{eqnarray*}
where $z_{(1)},\ldots,z_{(n+m)}$ denotes the ordered observations of the combined samples, $\bm{z}=\left(x_1,\ldots,x_n,y_1,\ldots,y_m\right)$. Additional optimizations are employed, for instance, minimization of the repeated computations such as sums and the creation of the necessary sequences, and the use of fast functions to sample the permutations and sort the vectors.

Two additional optimizations are: a) the fact that the combined samples are sorted and b) that the total sum does not change. When permuting the data, one needs to sort the observations that fall into the first permuted sample, the rest of the observations that form the second permuted sample are already sorted. Further, instead of sorting the observations of the first permuted sample, sort their indices, since it is faster to sort integers than numeric numbers. The $\delta_2$ quantity involves calculating the sum of each permuted sample. Since the total sum does not change, one needs to compute only the sum of the observations that fall within the permuted sample, and subtract it from the sum of all observations to obtain the sum of the observations of the second permuted sample.  

For the case of multivariate distributions, the commands \texttt{Rfast::dista(x, result = "sum")} and \texttt{Rfast::Dist(x, result = "sum")} compute the first term ($\delta_1$) and the other two ($\delta_2$ and $\delta_3$), respectively. The resulting function is not faster than the available implementation in the \textsf{energy} package, but it is memory-saving. Note that the command \texttt{Rfast::Dist(x, result = "sum")} can facilitate the memory-efficient computation of the multivariate normality test \citep{szekely2023} as well. 

\section{The relevant commands in \textsf{estats}} \label{secrfast}
The relevant commands in the \textsf{estats} package are \texttt{edist()}, \texttt{dvar()}, \texttt{dcov()}, \texttt{dcor()}, \texttt{pdcor()}, and \texttt{eqdist.etest()}, and we demonstrate their usage with the following examples. First, we consider two matrices with a few rows and a few columns. 

\subsection{The command \texttt{edist()}}
The command \texttt{edist()} accepts two numerical matrices as its arguments and computes the energy distance (\ref{energy}) between the two data sets.
\begin{verbatim}
x <- as.matrix( iris[1:50, 1:4] )
y <- as.matrix( iris[51:100, 1:4] )

estats::edist(x, y)
[1] 123.5538
\end{verbatim}

To compute the energy distance matrix among three or more data sets the user should assign them in a list and provide the list as a single argument in the function.
\begin{verbatim}
z <- as.matrix(iris[101:150, 1:4])
a <- list()
a[[ 1 ]] <- x
a[[ 2 ]] <- y
a[[ 3 ]] <- z

estats::edist(a)
         [,1]      [,2]      [,3]
[1,]   0.0000 123.55381 195.30396
[2,] 123.5538   0.00000  38.85415
[3,] 195.3040  38.85415   0.00000
\end{verbatim}

\subsection{The command \texttt{dvar()}}
The command \texttt{dvar()} accepts a numerical matrix and an extra logical argument (\textit{bc}) whose default is \textit{FALSE} and defines whether the bias-corrected distance variance (\ref{bcddvar}) should be computed or not.
\begin{verbatim}
estats::dvar(x)
[1] 0.2712927

estats::dvar(x, bc = TRUE)
[1] 0.06524269
\end{verbatim}

\subsection{The command \texttt{dcov()}}
The command \texttt{dcov()} accepts two numerical matrices and the logical argument (\textit{bc}) whose default is \textit{FALSE} and defines whether the bias-corrected distance covariance (\ref{bcdcov}) should be returned or not.
\begin{verbatim}
estats::dcov(x, y)
[1] 0.1025087

estats::dcov(x, y, bc = TRUE)
[1] -0.002748351
\end{verbatim}

\subsection{The command \texttt{dcor()}}
The command \texttt{dcor()} works in the same way as the command \texttt{dcov()} but returns a list with the (bias-corrected) distance variance of each data set, and their distance variance and correlation (\ref{dcor}).
\begin{verbatim}
estats::dcor(x,y)
 dcov     dvarX     dvarY      dcor 
0.1025087 0.2712927 0.4135274 0.3060479 

estats::dcor(x, y, bc = TRUE)
  dcov        dvarX        dvarY         dcor 
-0.002748351  0.065242693  0.156821104 -0.027170902 
\end{verbatim}

\subsection{The command \texttt{pdcor()}}
The command \texttt{pdcor()} unlike \texttt{dcor()} returns only the unbiased partial distance correlation.

\begin{verbatim}
estats::pdcor(x, y, z)
[1] -0.02722611
\end{verbatim}

\subsection{The command \texttt{eqdist.etest()}}
The command \texttt{eqdist.etest()} accepts two numerical vectors, or matrices, as its arguments and performs the energy test of equal univariate (or multivariate) distributions with the p-value computed via permutations. Even though the multivariate case is not fast enough, unlike the implementation in the \textsf{energy} package it is memory-efficient.
\begin{verbatim}
x <- iris[1:50, 1]
y <- iris[51:100, 2]

estats::eqdist.etest(x, y)
[1] 0.001
\end{verbatim}

\section{Measuring the computational cost and speed-up factors} \label{seccost}
This section illustrates the time improvements comparing the functions \texttt{dcor()} and \texttt{edist()} from the packages \textsf{energy}, \textsf{dcortools} and \textsf{estats}. The experiments were conducted utilising a Dell laptop equipped with Intel Core i5-1053G1 CPU at 1GHz, with 256 GB SSD, 8 GB RAM and Windows installed. Using a range of sample sizes $(n=(1,000, 2,000, 5,000, 10,000, 20,000, 50,000))$ and dimensions $(p=(2, 5, 10, 20))$ we generate two random matrices, and compute the running time, measured using the built-in command \textit{system.time()}, each function requires to compute the aforementioned quantities. 

The distance correlation examples were repeated, however, in this case we compared the running time required by the function \texttt{distcor()} that is available in the package \textsf{dcortools} with the option to use the memory save algorithm that according to the authors of the package has a computational complexity of $O(n^2)$ but requires only $O(n)$ memory.

\subsection{Results}
Tables \ref{times} and \ref{pdcortimes} present the running times of the implementations in \textsf{energy} and \textsf{estats} packages required for the computation of the distance correlation, energy distance, and partial distance correlation. Notably, \textsf{estats}'s implementation of the partial distance correlation is partially in \textit{R}. We note that for the case of sample sizes $n=20,000$ and higher, the \textsf{energy} functions produced the following error message "\textit{Error in .dcov(x, y, index) : 'R\_Calloc' could not allocate memory (20000 of 8 bytes)}".

Table \ref{trend} presents the estimated scalability of the \texttt{estats::dcor()} with regards to the sample size. The running time of the computation of both the distance correlation and the energy distance via \textsf{estats} increases quadratically with respect to the sample size. Note that the same rate was estimated for the function \texttt{dcortools::distcor()}. 

Finally, Table \ref{eqdist} contains the running time to perform the test of equality of two univariate distributions when the p-value is computed based on 999 permutations. Evidently, memory constraints preclude application of the test with tens of thousands of observations, but an additional constraint arises, that of computational time. In the case of 15,000 observations, the \texttt{energy::eqdist.etest()} requires prohibitively long computation time, which is more than a single day. In contrast, the \textit{R} function \texttt{estats::eqdist.etest()} requires 11 seconds for the case of 50,000 observations. Hence, the speed-up factors were not computed for this test.  

\begin{figure}[!ht]
\centering
\begin{tabular}{cc}
\includegraphics[scale = 0.45, trim = 0 0 0 0]{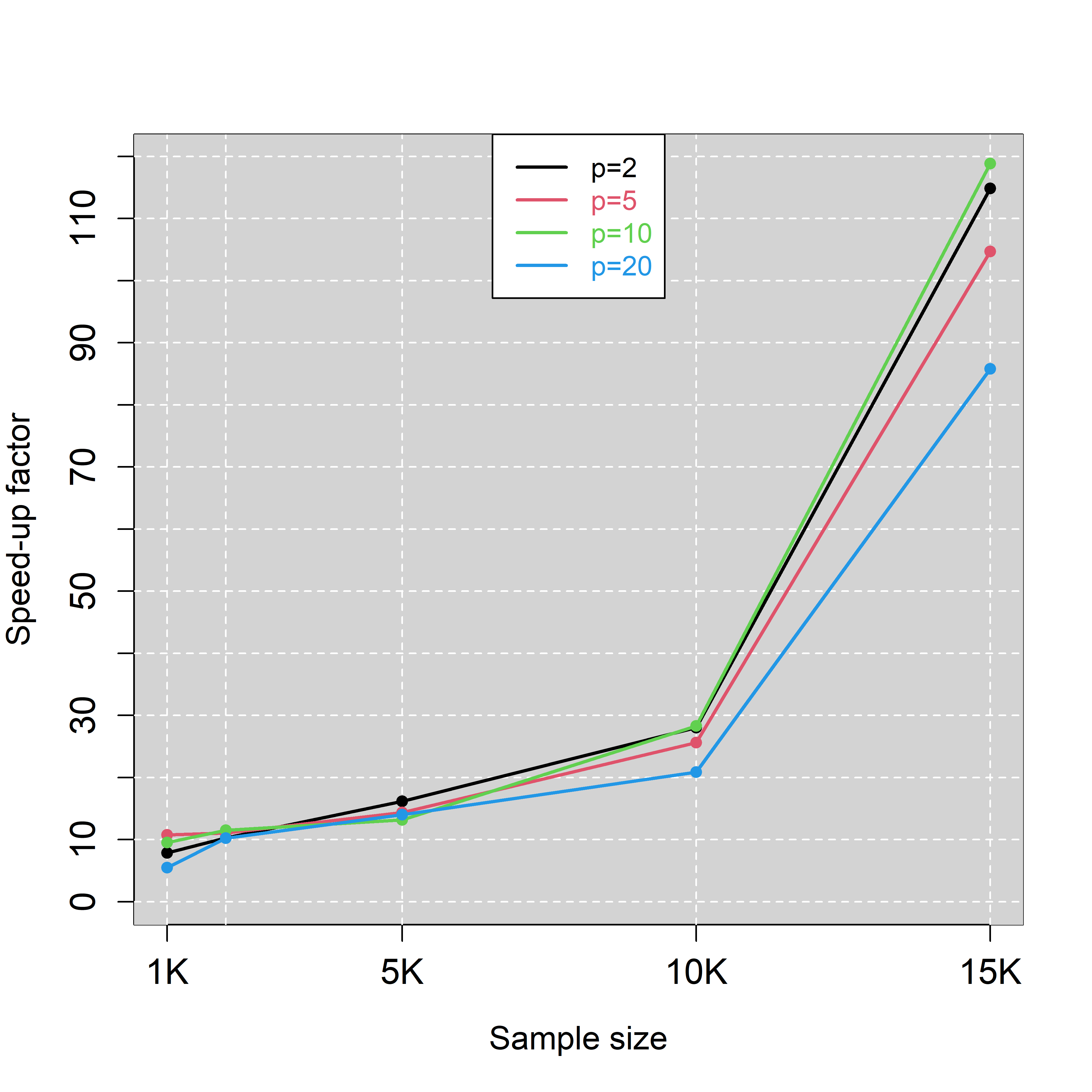} &
\includegraphics[scale = 0.45, trim = 30 0 0 0]{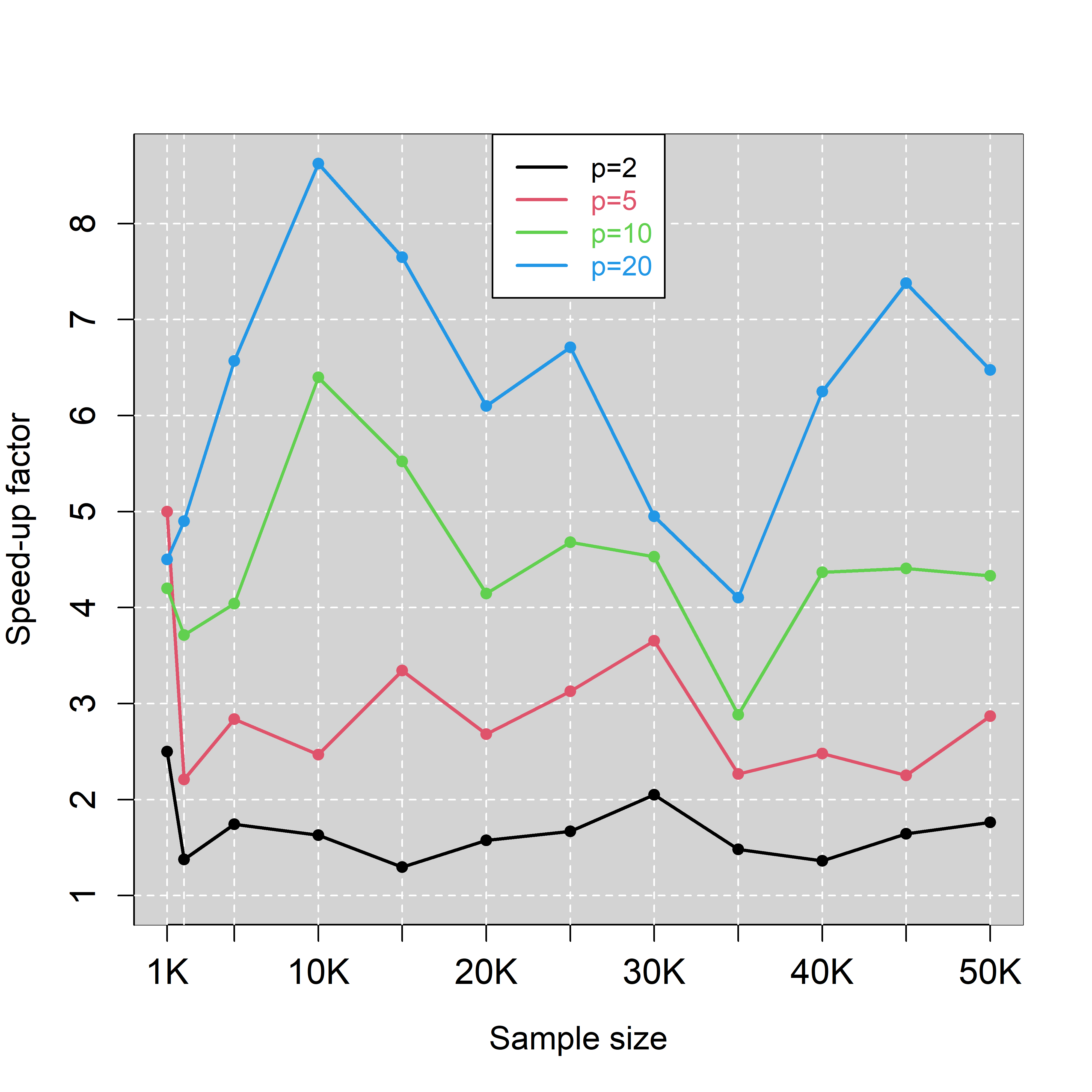}  \\
(a) \texttt{estats::dcor()} to \texttt{energy::dcor()}   &  (b) \texttt{estats::dcor()} to \texttt{dcortools::distcor()}  \\
\includegraphics[scale = 0.45]{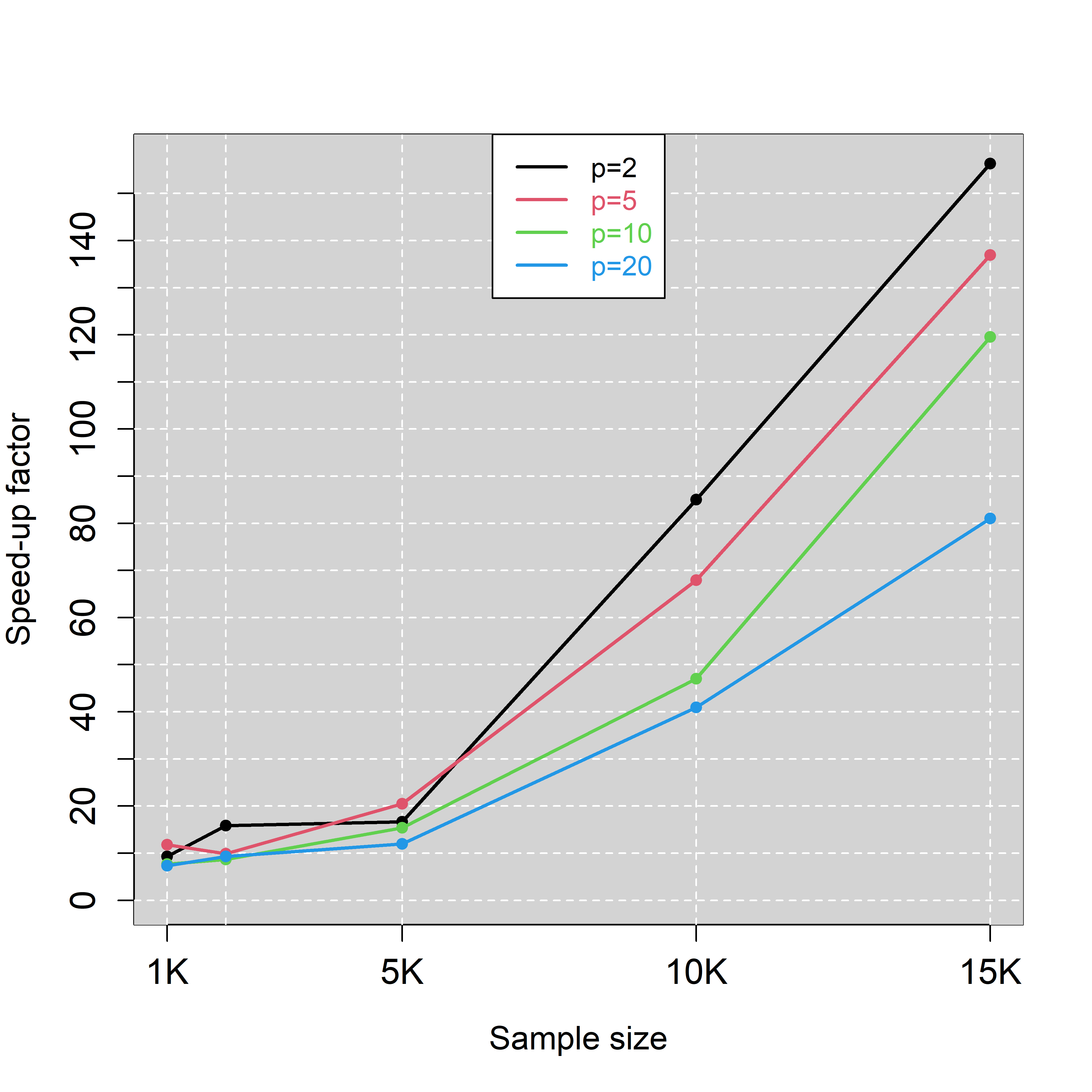}   &
\includegraphics[scale = 0.45]{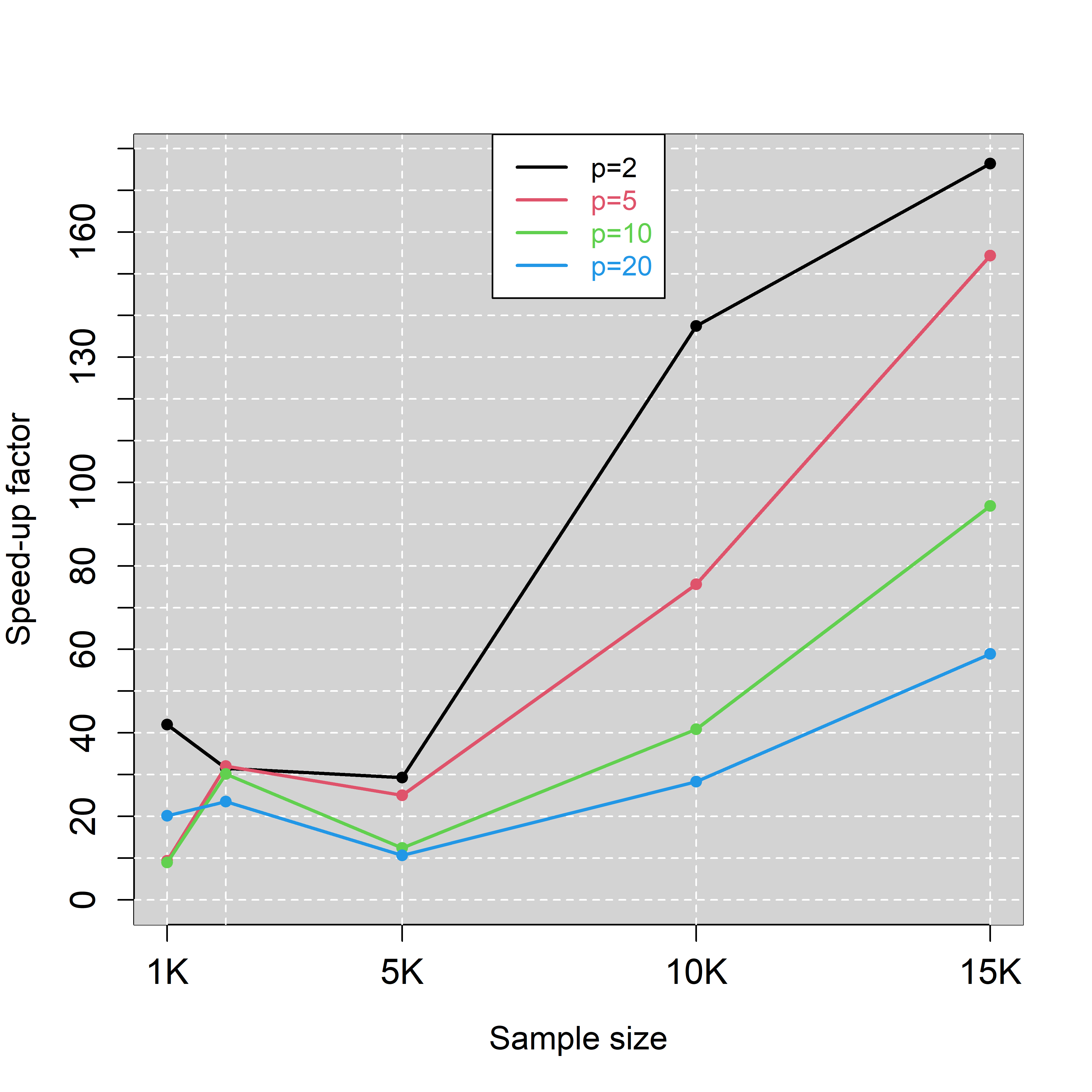}  \\
(c) \texttt{estats::pdcor()} to \texttt{energy::pdcor()}  & (d)  \texttt{estats::edist()} to \texttt{energy::edist()} 
\end{tabular}
\caption{Speed-up factors versus sample size for the \textsf{estats} functions of (a) and (b) distance correlation (c) partial distance correlation and (d) energy distance.}
\label{fig}
\end{figure}

Figure \ref{fig} presents the speed-up factors (ratio of the running time required by the functions in the package \texttt{energy} or \textsf{dcortools} divided by the running time required by the functions in the package \textsf{estats}). For the distance correlation, the speed-up factor ranges from 5.5 (for the case of $n=1,000$ and $p=20$) to 119 (for the case of $n=15,000$ and $p=10$) with an average of 33. Compared to \textsf{dcortools}, the \textsf{estats} function is slightly faster when $p=2$, and becomes faster (up to nearly 8 times faster) as the dimensions increase. For the partial distance correlation, the speed-up factor ranges from 7.33 (for the case of $n=1,000$ and $p=20$) up to 156 (for the case of $n=15,000$ and $p=2$). Finally, for the energy distance, the speed-up factor ranges from 4.6 (for the case of $n=1,000$ and $p=10$) and reaches 88 (for the case of $n=15,000$ and $p=2$) with an average of 26. 

The implementation of the distance correlation in the \textsf{energy} package, up to 15,000 observations, can be more than 100 times slower than our implementation, while the implementation in the \textsf{dcortools} package is on par or up to 8 times slower than our implementation, regardless of the sample size. The computational cost of the implementation in the \textsf{energy} package depends heavily on the sample size, while the implementation in the \textsf{dcortools} package becomes slower with increasing dimensionality. Similar patterns are observed for the speed-up factors of the partial distance correlation (Figure \ref{fig}(c)) and energy distance (Figure \ref{fig}(d)).

\begin{figure}[!ht]
\centering
\begin{tabular}{cc}
\includegraphics[scale = 0.45, trim = 0 0 0 0]{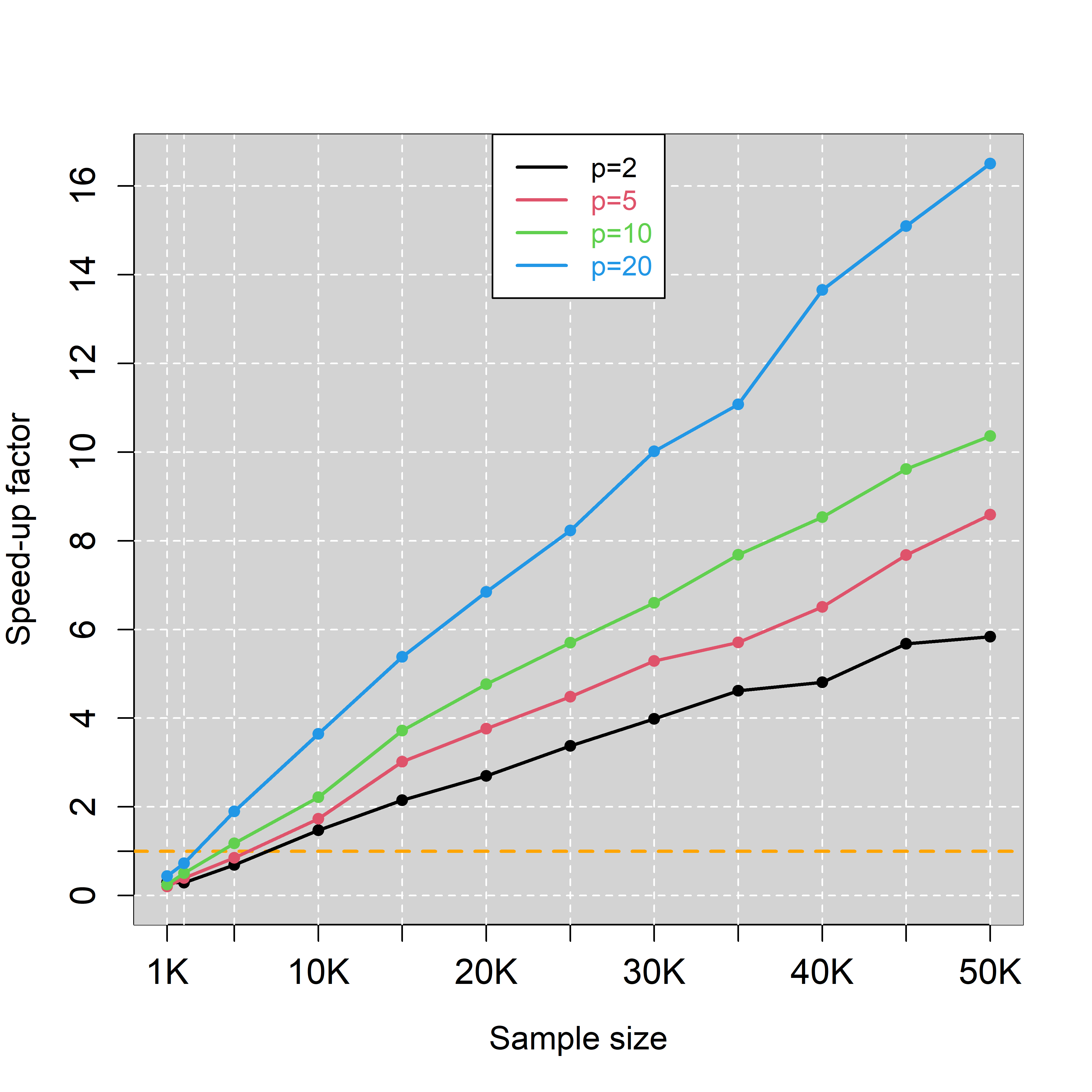} &
\includegraphics[scale = 0.45, trim = 30 0 0 0]{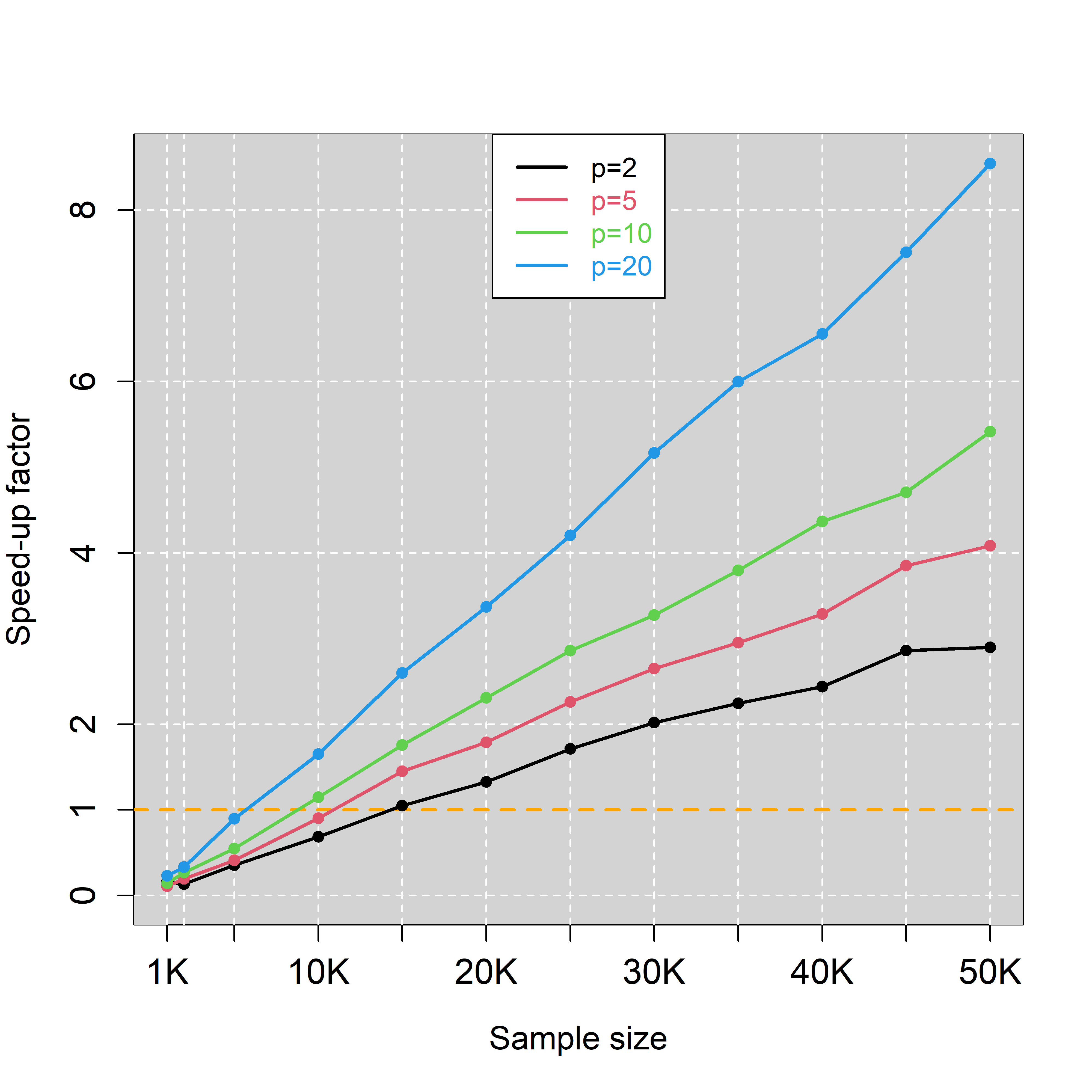}  \\
(a) \texttt{estats::dcov()} to \texttt{estats::adcov()}   &  (b) \texttt{estats::dcov()} to \texttt{estats::::adcov()}  \\
\includegraphics[scale = 0.45]{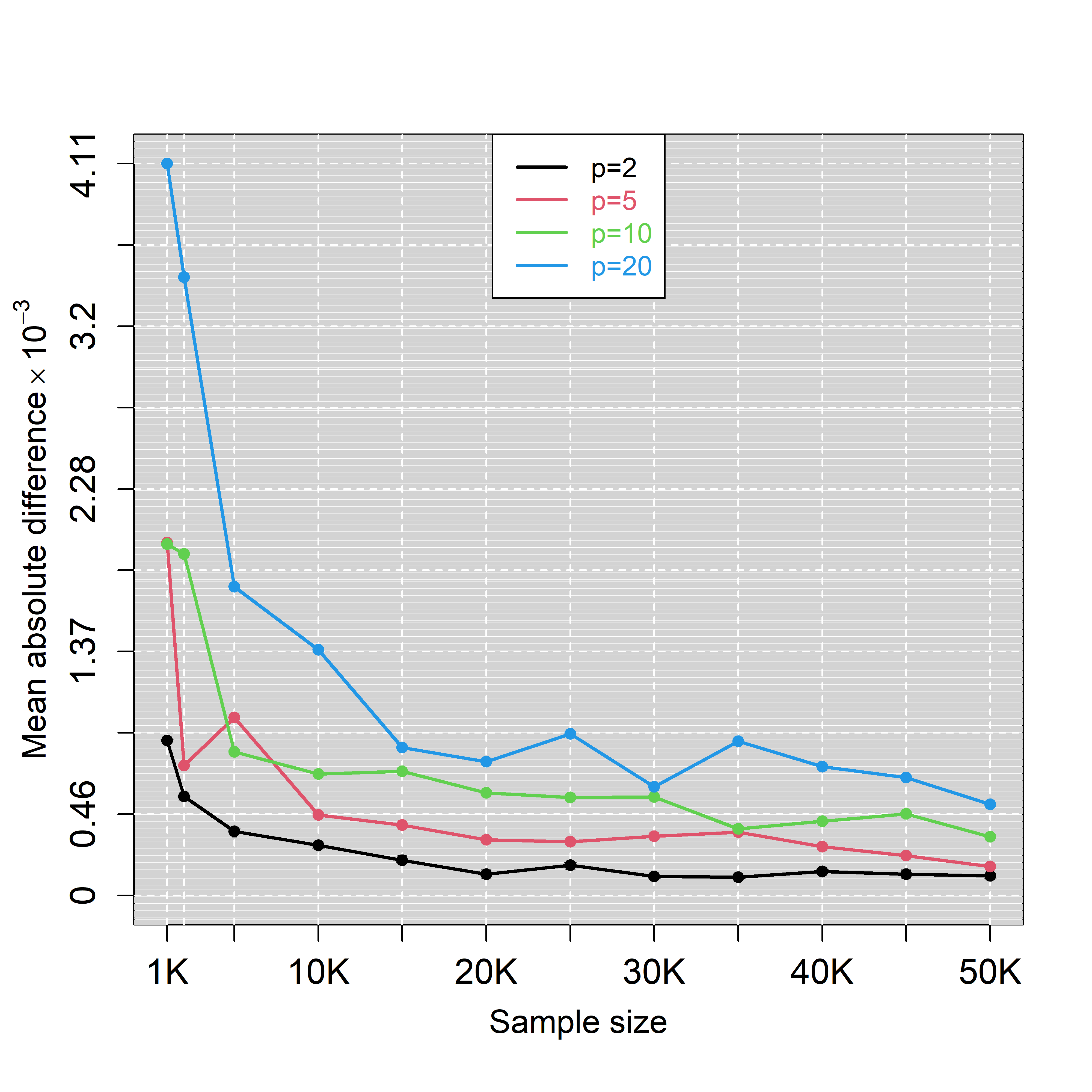}   &
\includegraphics[scale = 0.45]{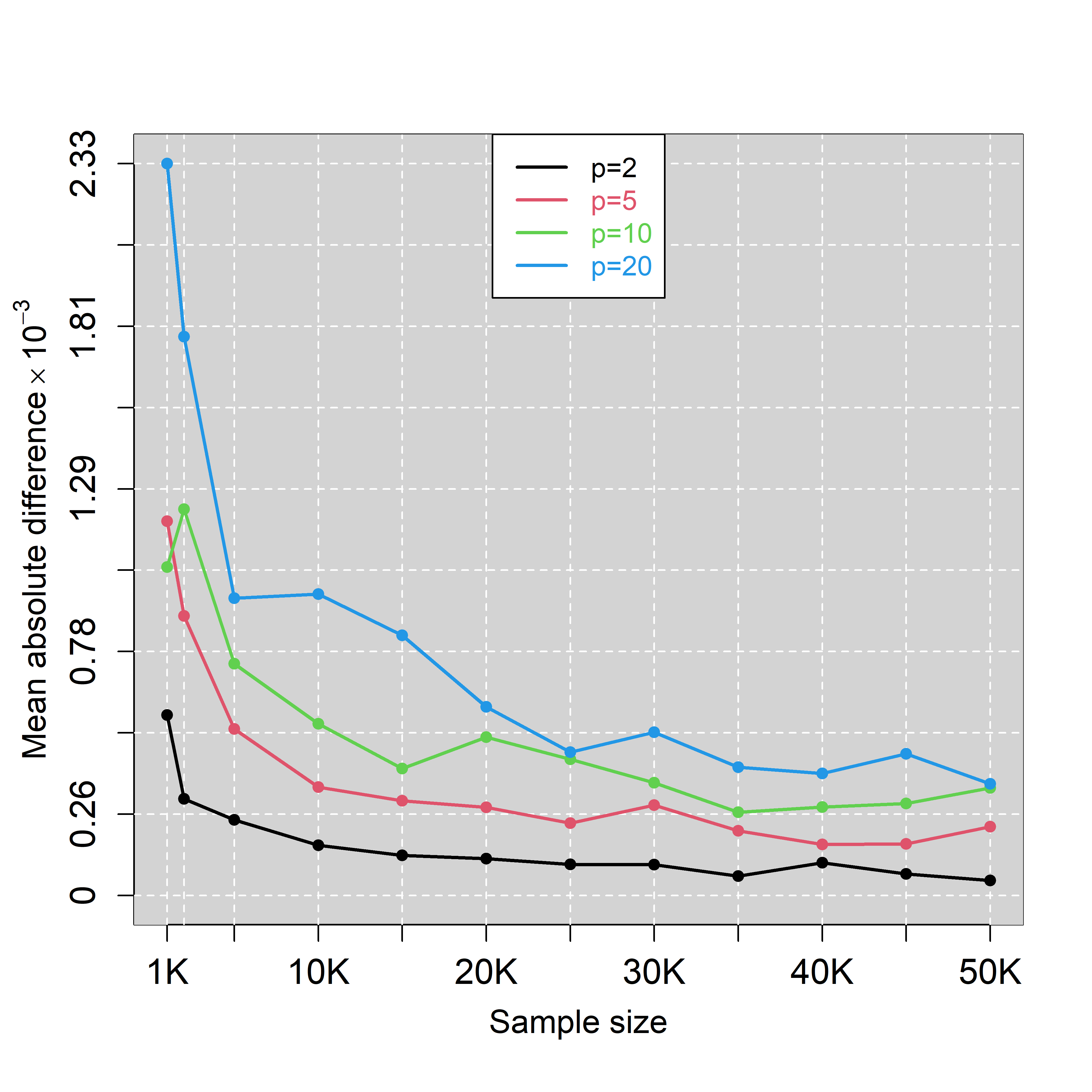}  \\
(c) \texttt{estats::adcov()} with K = 50  & (d)  \texttt{estats::adcov()} with K = 100  
\end{tabular}
\caption{Speed-up factors of the exact to the approximate distance covariance when (a) $K = 50$ and (b) $K = 100$. The mean absolute difference between the exact and the approximate distance covariance versus the sample size when (c) $K = 50$ and (d) $K = 100$.}
\label{fig2}
\end{figure}

Figure \ref{fig2} presents the speed-up factors and the mean absolute difference of the approximate distance covariance, using our implementation, available in the \textsf{estats} package, versus our implementation of the exact distance covariance. Figures \ref{fig2}(a) and \ref{fig2}(b) contain the speed-up factors when $K=50$ and $K=100$, respectively. Both figures demonstrate that computation of the exact distance covariance can be up to 16 times slower than the computation of the approximate distance covariance when $K=50$ and up to 8 times when $K=100$. In both cases, increasing the sample size increases the speed-up factor as well. Examination of Figures \ref{fig2}(c) and \ref{fig2}(d) reveals that larger values in $K$ yield higher accuracy. However, increasing the dimensionality of the matrices $\bm{X}$ and $\bm{Y}$ requires higher values of $K$. 

Note that we examined the case of independent multivariate variables and we cannot state an opinion on the accuracy of the approximation when the variables are dependent. This requires more research, but according to \cite{huang2022}, the approximate distance covariance is satisfactory. 

Collectively, for either quantity two key findings emerge. The speed-up factor increases with increasing sample size and second, the implementation in the package \textsf{energy} cannot compute those quantities when the sample size exceeds $15,000$ due to large memory requirements. Given the hardware specifications employed, this suggests that installation of a larger memory that would allow the quantities to be computed with the \textsf{energy} package, would yield greater performance improvements with increasing sample sizes.  

\begin{table}[ht]
\caption{Running times (in seconds) of the \textsf{energy} and the \textsf{estats}functions for the distance correlation and the energy distance. Dashes (–) denote that the quantity could not be computed by the \textsf{energy}'s function due to memory limitations.}
\label{times}
\centering
\begin{tabular}{l|l|rrrr|rrrr}
\toprule
&  & \multicolumn{4}{c}{Distance correlation} & \multicolumn{4}{c}{Energy distance} \\  \midrule
Sample size & Package & p=2 & p=5 & p=10 & p=20 & p=2 & p=5 & p=10 & p=20 \\  \midrule
$n=1,000$    & \textsf{energy} & 0.126 & 0.086 & 0.086 & 0.110 & 0.126 & 0.158 & 0.232 & 0.342 \\ 
            & \textsf{Rfast}  & 0.016 & 0.008 & 0.009 & 0.020 & 0.003 & 0.017 & 0.026 & 0.017 \\  \hline
$n=2,000$    & \textsf{energy} & 0.412 & 0.554 & 0.588 & 0.636 & 0.662 & 0.896 & 1.236 & 1.906 \\ 
            & \textsf{estats}  & 0.040 & 0.050 & 0.051 & 0.062 & 0.021 & 0.028 & 0.041 & 0.081 \\  \hline 
$n=5,000$    & \textsf{energy} & 4.446 & 4.424 & 4.512 & 5.670 & 4.012 & 4.300 & 5.258 & 7.750 \\ 
            & \textsf{estats}  & 0.275 & 0.308 & 0.342 & 0.404 & 0.137 & 0.172 & 0.424 & 0.731 \\  \hline 
$n=10,000$   & \textsf{energy} & 32.706 & 34.166 & 41.676 & 36.850 & 73.970 & 60.334 & 74.510 & 82.416 \\ 
            & \textsf{estats}  & 1.166 & 1.334 & 1.472 & 1.764 & 0.538 & 0.798 & 1.825 & 2.912 \\  \hline 
$n=15,000$   & \textsf{energy} & 321.894 & 336.263 & 410.177 & 362.679 & 353.902 & 365.658 & 389.848 & 395.272 \\ 
            & \textsf{estats}  & 2.803 & 3.212 & 3.451 & 4.227 & 2.006 & 2.369 & 4.130 & 6.710 \\  \midrule 
$n=20,000$   & \textsf{energy} & - & - & - & - & - & - & - & - \\ 
            & \textsf{estats}  & 5.079 & 6.008 & 6.189 & 7.540 & 3.289 & 4.330 & 6.984 & 1.954 \\  \hline 
$n=25,000$   & \textsf{energy} & - & - & - & - & - & - & - & - \\ 
            & \textsf{estats}  & 7.911 & 9.324 & 10.371 & 14.227 & 5.072 & 6.599 & 10.209 & 18.004 \\  \hline 
$n=30,000$   & \textsf{energy} & - & - & - & - & - & - & - & - \\ 
            & \textsf{estats}  & 13.681 & 13.770 & 20.622 & 24.061 & 7.152 & 9.425 & 16.454 & 29.571 \\  \hline 
$n=35,000$   & \textsf{energy} & - & - & - & - & - & - & - & - \\ 
            & \textsf{estats}  & 15.481 & 17.854 & 20.228 & 24.587 & 10.422 & 13.830 & 23.301 & 34.683 \\  \hline 
$n=40,000$   & \textsf{energy} & - & - & - & - & - & - & - & - \\ 
            & \textsf{estats}  & 21.725 & 24.420 & 26.223 & 32.256 & 13.611 & 17.417 & 25.143 & 56.997 \\  \hline 
$n=45,000$   & \textsf{energy} & - & - & - & - & - & - & - & - \\ 
            & \textsf{estats}  & 27.491 & 31.672 & 33.408 & 41.153 & 14.088 & 17.775 & 32.716 & 131.637 \\  \hline 
$n=50,000$   & \textsf{energy} & - & - & - & - & - & - & - & - \\ 
            & \textsf{estats}  & 33.959 & 38.181 & 40.899 & 52.201 & 17.105 & 22.413 & 49.346 & 223.218 \\ \bottomrule
\end{tabular}
\end{table}

\begin{table}[ht]
\caption{Running times (in seconds) of the \textsf{energy} and the \textsf{estats} implementations of the partial distance correlation. Dashes (–) denote that the quantity could not be computed by the \textsf{energy}'s function due to memory limitations.}
\label{pdcortimes}
\centering
\begin{tabular}{l|l|rrrr}
\toprule
&  & \multicolumn{4}{c}{Partial Distance correlation} \\  \midrule
Sample size & Package & p=2 & p=5 & p=10 & p=20 \\  \midrule
$n=1,000$  & \textsf{energy} & 0.242 & 0.354 & 0.246 & 0.264 \\ 
          & \textsf{estats}  & 0.026 & 0.030 & 0.032 & 0.036 \\ \hline
$n=2,000$  & \textsf{energy} & 1.458 & 0.970 & 1.106 & 1.728 \\ 
          & \textsf{estats}  & 0.092 & 0.098 & 0.128 & 0.186 \\ \hline 
$n=5,000$  & \textsf{energy} & 11.706 & 12.918 & 12.170 & 13.390 \\ 
          & \textsf{estats}  & 0.702 & 0.630 & 0.792 & 1.120 \\ \hline 
$n=10,000$ & \textsf{energy} & 264.614 & 235.116 & 238.410 & 277.240 \\ 
          & \textsf{estats}  & 3.112 & 3.460 & 5.074 & 6.774 \\ \hline 
$n=15,000$ & \textsf{energy} & 1321.772 & 1420.980 & 1436.202 & 1271.894 \\ 
          & \textsf{estats}  & 8.454 & 10.378 & 12.014 & 15.702 \\ \hline 
$n=20,000$ & \textsf{energy} & - & - & - & - \\ 
          & \textsf{estats}  & 15.050 & 14.788 & 19.224 & 31.688 \\ \hline 
$n=25,000$ & \textsf{energy} & - & - & - & - \\ 
          & \textsf{estats}  & 25.528 & 28.576 & 33.946 & 49.212 \\ \hline 
$n=30,000$ & \textsf{energy} & - & - & - & - \\ 
          & \textsf{estats}  & 33.562 & 34.834 & 50.248 & 70.294 \\ \hline 
$n=35,000$ & \textsf{energy} & - & - & - & - \\ 
          & \textsf{estats}  & 48.480 & 53.442 & 73.054 & 97.498 \\ \hline 
$n=40,000$ & \textsf{energy} & - & - & - & - \\ 
          & \textsf{estats}  & 61.976 & 69.916 & 92.764 & 113.812 \\ \hline 
$n=45,000$ & \textsf{energy} & - & - & - & - \\ 
          & \textsf{estats}  & 66.192 & 69.940 & 85.028 & 117.706 \\ \hline 
$n=50,000$ & \textsf{energy} & - & - & - & - \\ 
          & \textsf{estats}  & 77.248 & 85.526 & 108.740 & 146.986 \\ \hline 
 \bottomrule
\end{tabular}
\end{table}

\begin{table}[!ht]
\caption{Estimated scalability rate of the \textsf{estats} functions for the distance correlation, partial distance correlation and the energy distance (95\% confidence interval within parentheses).}
\label{trend}
\centering
\begin{tabular}{l|cccc}
\toprule
Case           & p=2 & p=5 & p=10 & p=20 \\  \midrule
\texttt{dcor()}  & 2.022 (1.952, 2.092) & 2.131 (2.085, 2.177) & 2.143 (2.078, 2.208) & 2.062 (1.995, 2.130) \\
\texttt{edist()} & 2.210 (2.117, 2.303) & 1.987 (1.852, 2.122) & 2.002 (1.885, 2.119) & 2.265 (2.101, 2.429) \\ 
\texttt{pdcor()} & 2.113 (2.050, 2.177) & 2.117 (2.031, 2.203) & 2.145 (2.060, 2.229) & 2.164 (2.076, 2.253) \\ \bottomrule
\end{tabular}
\end{table}

\begin{table}[!ht]
\caption{Running times (in seconds) of the \textsf{energy} and \textsf{estats} functions for the equality of univariate distributions with $999$ permutations. Dashes (–) denote that the quantity could not be computed by the \textsf{energy} package function due to memory limitations. \\
$^*$In the case of $n=15,000$ the \textsf{energy} implementation required \textit{516.082} seconds for only 4 permutations. Consequently, completing 999 permutations would require more than 24 hours of computation time.}
\label{eqdist}
\centering
\begin{tabular}{l|rr}
\toprule 
Sample size & \textsf{energy} & \textsf{estats} \\ \midrule 
$n=1,000$  & 5.068   & 0.138 \\ 
$n=2,000$  & 28.512  & 0.320 \\ 
$n=5,000$  & 178.040 & 0.782 \\ 
$n=10,000$ & 712.880 & 1.940 \\ 
$n=15,000$ & \textit{516.082}$^* $ & 3.240 \\ 
$n=20,000$ & - & 4.172 \\ 
$n=25,000$ & - & 4.734 \\ 
$n=30,000$ & - & 5.598 \\ 
$n=35,000$ & - & 7.186 \\ 
$n=40,000$ & - & 6.038 \\ 
$n=45,000$ & - & 6.706 \\ 
$n=50,000$ & - & 7.900 \\  
\bottomrule 
\end{tabular}
\end{table}

\clearpage
\section{Discussion} \label{secconc}
We presented several algorithmic improvements to reduce the memory and computational cost associated with the energy distance, the distance variance, covariance and (partial) correlation measures, and an equality of univariate distributions test. These improvements leverage on memory-efficient computations of distance related sums, and involve mathematical identities. We compared our functions to existing implementations in other \textit{R} packages, such as \textsf{energy} and \textsf{dcortools} and we demonstrated that our implementations achieve substantially faster execution times, and in contrast to the functions available in \textsf{energy}, ours are applicable to large datasets, since they require small amounts of memory and can be used even on computing systems with limited resources. Note that the functions in \textsf{estats} and \textsf{dcortools} use a combination of \textit{R} and \textit{C++}, while the functions in \textsf{energy} use a combination of \textit{R} and \textit{C}. 

The key improvements of our implementations are: a) we reduced the memory requirements from $O(n^2)$ to $O(n)$, b) we achieved $5-156 \times$ speedup over existing implementations, c) we enabled the analysis of datasets with tens of thousands of observations faster than any other implementation and d) we provide exact and not approximate results. e) we investigated the approximate distance covariance and observed that with large-scale data, trading accuracy for computational efficiency may be advantageous.

Regarding computations with big data, the \textit{R} package \textsf{bigmemory} \citep{bigmemory2013} is widely adopted. However, a limitation of this package is that if the data can be loaded in \textit{R} they must first be converted to the \textit{big.matrix} class prior to using that package. Additionally, computations would be slower than the ones we have already implemented in \textit{C++}. Thus, this would increase the computational complexity and our purpose is to not only compute the required statistics (distance correlation, covariance, variance, etc.) without using excessive memory, but highly efficient as well.

These techniques extend to the computation of $U$--statistics related quantities \citep{lee2019} and in kernel density estimation \citep{wand1994}. Kernel functions are used in the context of maximum mean discrepancy hypothesis testing \citep{gretton2012} and hence can be accelerated and become memory-efficient. Another application of kernel density estimation, which has important applications in economics, is the computation of the income polarization index \citep{esteban2004}, even with hundreds of thousands of observations. The computational efficiency of the Nadaraya-Watson regression \citep{wand1994} can also benefit from these optimizations, as in the case of the kernel regression for compositional data \citep{tsagris2023}.

We have applied these techniques to kernel density in \textsf{Rfast2} and for the computation of the polarization index of \cite{esteban2004} in the package \textsf{DER} \citep{der2025}, but further extensions are possible. Future work includes implementing the hypothesis testing of equality of univariate distributions in \textit{C++} and allow for parallel computations of the distances, a key improvement that will reduce the computation time for all the $\mathcal{\varepsilon}$--statistics quantities presented here.

\bibliographystyle{apalike}
\bibliography{vivlio}

\end{document}